\documentclass[preprint,12pt]{aastex}
\newlength{\GBCdigit}
\newcommand{\GBC}{\hspace*{\GBCdigit}}
\newlength{\GBCminus}
\newcommand{\Gbc}{\hspace*{\GBCminus}}

\shorttitle{HETE-2 Observation of GRB 021211}
\shortauthors{Crew et al.}

\begin{document}

\title{HETE-2 Localization and Observation of
the Bright, X-Ray-Rich Gamma-Ray Burst GRB 021211}

\author{
G.~B.~Crew,\altaffilmark{1}
D.~Q.~Lamb,\altaffilmark{2}
G.~R.~Ricker,\altaffilmark{1}
J.-L.~Atteia,\altaffilmark{3}
N.~Kawai,\altaffilmark{4,5}
R.~Vanderspek,\altaffilmark{1}
J.~Villasenor,\altaffilmark{1}
J.~Doty,\altaffilmark{1}
G.~Prigozhin,\altaffilmark{1}
J.~G.~Jernigan,\altaffilmark{6}
C.~Graziani,\altaffilmark{2}
Y.~Shirasaki,\altaffilmark{5,7}
T.~Sakamoto,\altaffilmark{4,5,8}
M.~Suzuki,\altaffilmark{4}
N.~Butler,\altaffilmark{1}
K.~Hurley,\altaffilmark{6}
T.~Tamagawa,\altaffilmark{5}
A.~Yoshida,\altaffilmark{5,9}
M.~Matsuoka,\altaffilmark{10}
E.~E.~Fenimore,\altaffilmark{8}
M.~Galassi,\altaffilmark{8}
C.~Barraud,\altaffilmark{3}
M.~Boer,\altaffilmark{11}
J.-P.~Dezalay,\altaffilmark{11}
J.-F.~Olive,\altaffilmark{11}
A.~Levine,\altaffilmark{1}
G.~Monnelly,\altaffilmark{1}
F.~Martel,\altaffilmark{1}
E.~Morgan,\altaffilmark{1}
T.~Q.~Donaghy,\altaffilmark{2}
K.~Torii,\altaffilmark{5}
S.~E.~Woosley,\altaffilmark{12}
T.~Cline,\altaffilmark{13}
J.~Braga,\altaffilmark{14}
R.~Manchanda,\altaffilmark{15}
G.~Pizzichini,\altaffilmark{16}
K.~Takagishi,\altaffilmark{17}
~and~M.~Yamauchi\altaffilmark{17}
}

\altaffiltext{1}{Center for Space Research, Massachusetts Institute of
Technology, 70 Vassar Street, Cambridge, MA, 02139.}

\altaffiltext{2}{Department of Astronomy and Astrophysics, University
of Chicago, 5640 South Ellis Avenue, Chicago, IL 60637.}

\altaffiltext{3}{Laboratoire d'Astrophysique, Observatoire
Midi-Pyr\'{e}n\'{e}es, 14 Ave. E. Belin, 31400 Toulouse, France.}

\altaffiltext{4}{Department of Physics, Tokyo Institute of Technology, 
2-12-1 Ookayama, Meguro-ku, Tokyo 152-8551, Japan.}

\altaffiltext{5}{RIKEN (Institute of Physical and Chemical Research),
2-1 Hirosawa, Wako, Saitama 351-0198, Japan.}

\altaffiltext{6}{University of California at Berkeley,
Space Sciences Laboratory, Berkeley, CA, 94720-7450.}

\altaffiltext{7}{National Astronomical Observatory, Osawa 2-21-1,
Mitaka,  Tokyo 181-8588 Japan.}

\altaffiltext{8}{Los Alamos National Laboratory, P.O. Box 1663, Los 
Alamos, NM, 87545.}

\altaffiltext{9}{Department of Physics, Aoyama Gakuin University,
Chitosedai 6-16-1 Setagaya-ku, Tokyo 157-8572, Japan.}

\altaffiltext{10}{Tsukuba Space Center, National Space Development
Agency of Japan, Tsukuba, Ibaraki, 305-8505, Japan.}

\altaffiltext{11}{Centre d'Etude Spatiale des Rayonnements,
Observatoire Midi-Pyr\'{e}n\'{e}es,
9 Ave. de Colonel Roche, 31028 Toulouse Cedex 4, France.}

\altaffiltext{12}{Department of Astronomy and Astrophysics, University 
of California at Santa Cruz, 477 Clark Kerr Hall, Santa Cruz, CA
95064.}

\altaffiltext{13}{NASA Goddard Space Flight Center, Greenbelt, MD,
20771.}

\altaffiltext{14}{Instituto Nacional de Pesquisas Espaciais, Avenida
Dos Astronautas 1758, S\~ao Jos\'e dos Campos 12227-010, Brazil.}

\altaffiltext{15}{Department of Astronomy and Astrophysics, Tata 
Institute of Fundamental Research, Homi Bhabha Road, Mumbai, 400 005, 
India.}

\altaffiltext{16}{Consiglio Nazionale delle Ricerche,
IASF, Sezione di Bologna, via Piero Gobetti 101, 40129 Bologna, Italy.}

\altaffiltext{17}{Faculty of engineering, Miyazaki University, Gakuen
Kibanadai Nishi, Miyazaki 889-2192, Japan.}

\begin{abstract}
A bright, X-ray-rich gamma-ray burst (GRB) was detected by the French
Gamma Telescope (FREGATE) and localized with the Wide-field X-ray
Monitor (WXM) and Soft X-ray Camera (SXC) instruments on the High
Energy Transient Explorer 2 satellite (HETE-2) at 11:18:34.03 UT
(40714.03 SOD) on 11 December 2002. The WXM flight software localized
the burst to a 14\arcmin\ radius; this was relayed to the astronomical
community 22 seconds after the start of the burst. Ground analysis of
WXM and SXC data provided refined localizations; the latter can be
described as a circle with a radius of 2\arcmin\, centered at R.A.
08$^{\rm h}$ 09$^{\rm m}$ 00$^{\rm s}$, Dec  06\degr\ 44\arcmin\
20\arcsec\ (J2000).
GRB 021211 consists of a single, FRED-like pulse with a duration 
$t_{90} \approx 2.3$ s at high energies (85--400 keV) which increases
to $t_{90} \approx 8.5$ s at low energies (2--10 keV).  The peak photon
number and photon energy fluxes in the 2--400 keV band are
$(34.0 \pm 1.8)$ ph cm$^{-2}$ s$^{-1}$ and
$(1.68 \pm 0.11)\times 10^{-6}$ erg cm$^{-2}$ s$^{-1}$, respectively.
The energy fluences in the 2--30 keV and 30--400 keV energy bands are
$S_X = (1.36 \pm 0.05) \times 10^{-6}$ erg cm$^{-2}$ and 
$S_\gamma = (2.17 \pm 0.15) \times 10^{-6}$ erg cm$^{-2}$, 
respectively.  Thus GRB 021211 is an ``X-ray-rich'' GRB ($S_X/S_\gamma =
0.63 > 0.32$).
The average spectrum of the the burst is well-fit by a Band function
(low-energy power-law index $\alpha = -0.805^{+0.112}_{-0.105}$;
high-energy power-law index $\beta = -2.37^{+0.18}_{-0.31}$; and energy
of the peak of the spectrum in $\nu F_\nu$, $E^{\rm obs}_{\rm peak} =
46.8^{+5.8}_{-5.1}$ keV).
The near-real time optical follow-up of GRB 021211 made possible by
HETE-2 led to the detection of an optical afterglow for what otherwise
would quite likely have been classified as an ``optically dark'' GRB,
since the optical transient faded rapidly (from $R < 14$ to $R \approx
19$) within the first 20 minutes, and was fainter than $R \approx 23$
within 24 hours after the burst.  GRB 021211 demonstrates that some
fraction of burst afterglows are ``optically dark'' because their
optical afterglows at times $>$ 1 hour after the burst are very faint,
and previously have often escaped detection.  Such bursts are
``optically dim'' rather than truly ``optically dark.''  GRB 021211
also shows that even such ``optically dim'' bursts can  have very
bright optical afterglows at times $<$ 20 minutes after the burst.
\end{abstract}


\section{Introduction}

The bane of the gamma-ray burst (GRB) field are the so-called
``optically dark'' bursts:  those frustrating events for which, despite
the best efforts of the satellite observers to provide an accurate and
rapid location for the burst, and the diligence of the followup
community in searching for a corresponding transient event, in the end,
only upper limits on any optical afterglow can be established.  In this
paper we report on an event that provides a partial answer to the
puzzle of such ``optically dark'' GRBs.

On Dec 11 2002, at 11:18:34.03 UT (40714.03 SOD),
the High Energy Transient Explorer 2 Satellite
(HETE-2; Ricker et al. 2003)
detected a GRB 
\citep{gcn1734}, 
with the French Gamma Telescope (FREGATE) %
\citep{WH2001-FREGATE}. 
GRB 021211 lasted only a few seconds, but it was bright enough 
and soft enough to allow the
Wide-field X-ray Monitor (WXM) 
\citep{WH2001-WXM} 
to unambiguously locate its position.  The
Soft X-ray Camera (SXC) 
\citep{WH2001-SXC-jsv} 
also easily detected the burst.
The WXM flight location was relayed to the
ground via the burst alert network %
(Crew et al. 2003; Villasenor et al. 2003.; Vanderspek et al. 2003)
22 s after the start of the burst.  HETE-2 was at this moment emerging
from a no-contact zone over the Pacific Ocean---the burst message was
caught on the horizon by the HETE-2 secondary ground station in the
Galapagos Islands.  

Three robotic telescopes, RAPTOR (RAPid Telescopes for Optical
Response), KAIT (Katzman Automatic Imaging Telescope), and Super-LOTIS
(Livermore Optical Transient Imaging System), caught the optical
transient during the second minute of the event %
\citep{gcn1757,gcn1737,gcn1736}
and a fourth provided the identification of the optical afterglow only
54 minutes after the start of the burst \citep{gcn1731}.  However, the
afterglow had already faded to R $\approx$ 19 at the time of this
fourth observation, and subsequent observations showed that it had
faded to $>$ 23rd magnitude within 24 hours of the burst
\citep{gcn1739,gcn1744,gcn1750,fox03b,pandey2003}.
The host galaxy of the burst was identified in HST images taken
on 18 December %
\citep{gcn1781}; 
soon thereafter, the redshift of the host galaxy was determined to be
$z = 1.006$ %
\citep{gcn1785}. 

The optical afterglow of GRB 021211 is only the third to be caught in
near-real time, the other two being GRB 990123 %
\citep{briggs99,akerlof99}
and GRB 021004 \citep{shirasaki2002,doty2002,lamb2002,fox03a}.
The afterglow of GRB 021211 was more than three magnitudes fainter than
was the afterglow of GRB 990123 at the same epochs, and about 70\% of
the searches conducted to date would have failed to detect it at one
day after the burst (see Berger et al., Figure 4; Lazzati, Covino \&
Ghisellini 2002, Figure 1; Fox et al.  2003b, Figure 2).

We turn now to a discussion of the properties of the
prompt X-ray and $\gamma$-ray emission from GRB 021211, which will allow
a comparison of this event with GRB 990123 (and other bursts with bright
optical afterglows), in the hope that such a comparison may shed some
light on the nature of ``optically dark'' bursts.  We show that
GRB 021211, while having a duration that is shorter than that of most
long-duration bursts and lying near the boundary in duration between
long and short GRBs (Hurley 1992; Lamb, Graziani, and Smith 1993;
Kouveliotou et al. 1993)
is definitely a long burst, by virtue of its spectral properties.
Indeed, the burst is ``X-ray-rich,'' with a spectrum that peaks at
$E^{\rm obs}_{\rm peak} \approx 47$ keV.  The duration of the burst
decreases with increasing energy and exhibits strong spectral evolution
from hard to soft, which are also properties that are typical of long
GRBs.

\section{Observations} \label{observations}

\subsection{Localization} \label{localization}

At the time of the burst, the HETE-2 spacecraft was pointed $\approx$
30\degr\ away from its nominal anti-solar pointing direction, in order
to make calibration observations of the Crab. %
The burst triggered the FREGATE instrument on the 0.160 s trigger
timescale in the FREGATE 85--400 keV energy band.  The WXM flight
software determined an initial position using the first 3 s of data
from the burst; this was relayed to the ground in the {\it Alert}
message.  Several seconds later it found a better {\it Update} position
using the first 4.8~s of data after the trigger: 
R.A. = 08$^{\rm h}$ 08$^{\rm m}$ 54.7$^{\rm s}$, 
Dec = 06\degr\ 44\arcmin\ 05\arcsec. 
Both positions were assigned a 14\arcmin\ error radius based on the
strength of the image signal to noise ratios ($S/N = 29.2$ and 33.5 in
the X- and Y-detectors, respectively).
The WXM localization was also relayed to the flight SXC software, which
correctly localized the burst.  However, the SXC flight localization
was not sent out to the astronomical community because the calibration
of SXC flight localizations had not yet been completed.

\begin{figure}
\includegraphics[scale=0.9,clip=]{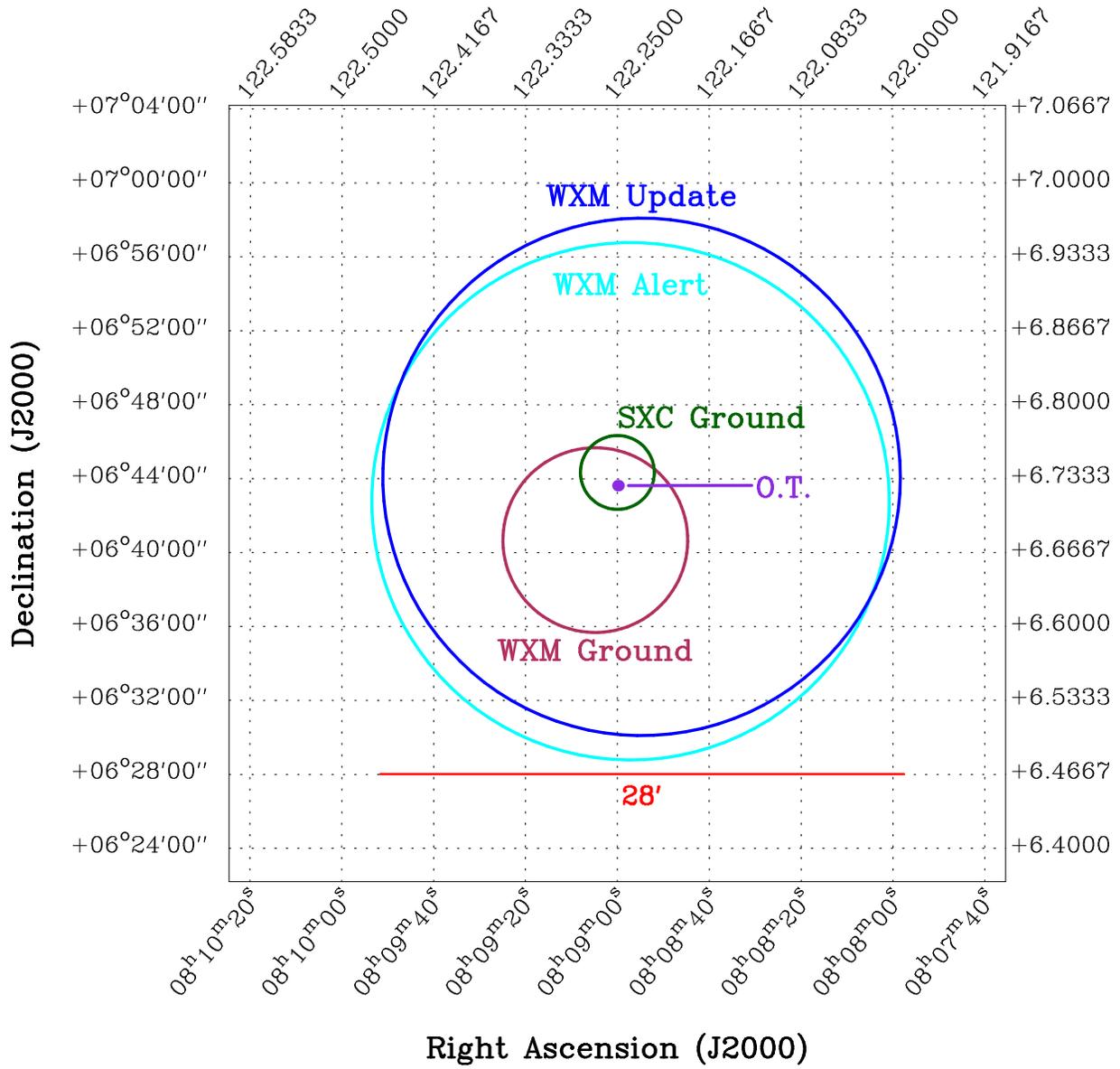}
\caption{Skymap summarizing the localizations reported in the
GCN Burst Position Notices (Seq. Num. 1--5); cf. Table \ref{tbl:location}.
The 28\arcmin\ diameter for the WXM Alert and Update flight positions
is nominal.
\label{fig:skymap}}
\end{figure}

Owing to preventive maintenance of the Cayenne Primary Ground Station, 
burst buffer data were not relayed to the ground until an hour after the 
burst.  The WXM and SXC data were then analyzed
\citep{carlo03,yuji03,WH2001-SXC-gm},
and produced refined localizations that are in good agreement with the
WXM flight localization (see Figure \ref{fig:skymap}).
Using a 4--5~s foreground
region (similar to the choice made by the flight software), the radius
of the 90\% confidence region for the burst location is $\approx$
3\arcmin\ from statistical error alone; including systematic errors, it
is conservatively 5\arcmin.  The SXC analysis was also straightforward.
The 2\arcmin\ radius of the final SXC localization, centered at R.A. =
08$^{\rm h}$ 09$^{\rm m}$ 00.0$^{\rm s}$, Dec  = 06\degr\ 44\arcmin\ 
20\arcsec, is dominated by systematic errors.  The localization history
is summarized in Table \ref{tbl:location}.

GRB 021211 was observed by other spacecraft in the Interplanetary
Network, and the resulting IPN localization \citep{hurley2002} is fully
consistent with the WXM and SXC localizations.  All of these
localizations are consistent with the location of the optical transient
\citep{gcn1731}, as shown in Table \ref{tbl:location}
and Fig. \ref{fig:skymap}.

\subsection{Temporal Properties} \label{time_history}

\begin{figure}
\includegraphics[scale=0.9,clip=]{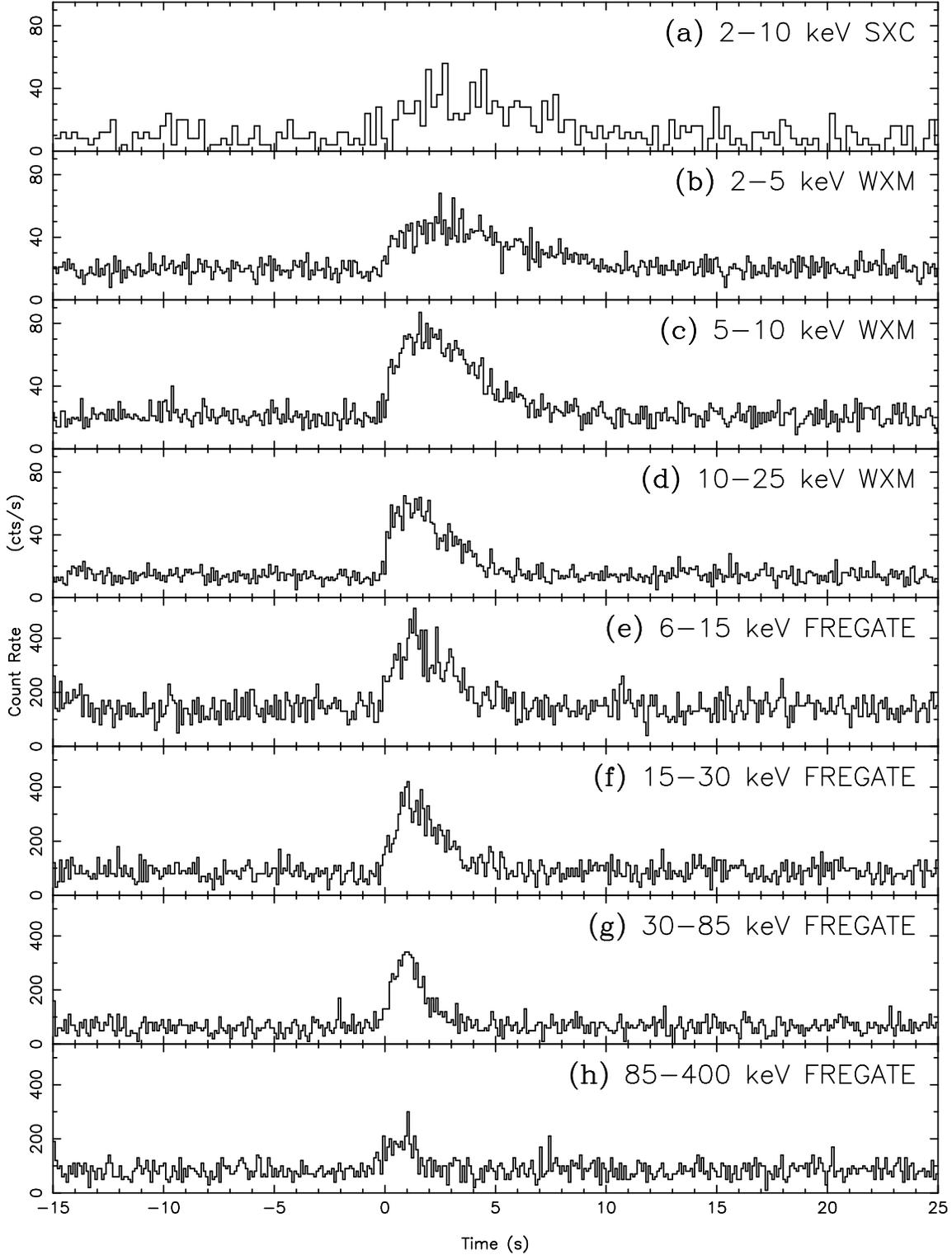}
\caption{Time history of GRB 021211 as observed by the HETE-2 instruments:
(a) all SXC data (2--10 keV) binned in 250 ms bins; WXM data binned in
100 ms bins and (b) 2--5 keV, (c) 5--10 keV, and (d) 10--25 keV bands;
and FREGATE data binned in 100 ms bins and (e) 6--15 keV, (f) 15--30
keV, (g) 30--85 keV, and (h) 85--400 keV bands.
\label{fig:sxc}
\label{fig:wxm}
\label{fig:fregate}}
\end{figure}

The burst time history is FRED-like (fast-rise, exponential decay) and
relatively featureless in all energy bands of all instruments. The
burst lasts $\sim$10~s in the SXC (2--10 keV) as shown in Fig.
\ref{fig:sxc}a.  The SXC data have been binned in 250 ms bins in order
to resolve the fast rise of the burst.  The duration of the burst in
the WXM decreases from $\approx 20$ s in the 2--5 keV energy band to
$\approx 4$ s in the 10--25 keV energy band, as shown in Fig.
\ref{fig:wxm}b--d; the data are binned in 100 ms bins.  Fig.
\ref{fig:fregate}e--h shows the time history of the burst in various
FREGATE energy bands, again binned in 100 ms bins.  The duration of the
burst decreases to $\approx 2.3$ s in the highest FREGATE energy band
(85--400 keV).

Table \ref{tbl:temporal} gives the $t_{50}$ and $t_{90}$ durations for
various SXC, WXM and FREGATE energy bands.  Figure 3 shows that both
duration measures decrease with increasing energy: $t_{50}$ = $4.2
E^{-0.34}$ (s/keV) and $t_{90}$ = $13 E^{-0.40}$ (s/keV).  This
behavior is typical of  long-duration GRBs 
\citep{fenimore95}. 
The SXC and WXM data both suggest a departure from this relation at the
lowest energies ($<$ 5 keV), and there is indeed some evidence in the
WXM 2--5 keV time history for a weak, soft tail (see 
Figures \ref{fig:wxm}b and 3).

\subsection{Spectrum} \label{spectrum}

We investigated the average spectral properties of the burst, using an
interval of 8.0~s, which corresponds to the $t_{90}$ of the burst in
the WXM 2--25 keV energy band (see Table \ref{tbl:temporal}). 
To examine the spectral evolution of the burst, we also analyzed the
spectra for two subintervals:  t = 0--2.3 s, in which the high energy
photons are clearly seen in the FREGATE 85--400 keV band in Fig.
\ref{fig:fregate}h; and t = 2.3--8.0 s, in which they are absent.

Table \ref{tbl:spectrum}
presents the results of our time-resolved and time-integrated
spectral analysis of the burst.  In this analysis, we consider two
models:  (1) power law times exponential (PLE) [the COMP model in
\cite{preece00}], and (2) Band function \citep{band93}.  Both the PLE
and the Band function models provide satisfactory descriptions of the
data: for the PLE model, $\chi^2_\nu = 0.78$ (111 DOF), while for the
Band function, $\chi^2_\nu = 0.68$ (110 DOF).  However, the difference
between $\chi^2_{\rm min}$ for the PLE and Band function models is
$\Delta \chi^2 = 11.57$ for one additional parameter.  Thus the Band
function is preferred at the $6.7 \times 10^{-4}$ confidence level.

Figure \ref{fig:spec-cts} shows the fit of the Band function to the WXM
and FREGATE spectral data.
The energy $E^{\rm obs}_{\rm peak} = (2 - \alpha) E_0$ of the peak of
the $\nu F_\nu $spectrum is $46.8^{+5.8}_{-5.1}$ keV, for the
burst-average spectrum, which is at the low end of the range of $E^{\rm
obs}_{\rm peak}$ values observed by BATSE
\citep{preece00} 
and BeppoSAX 
\citep{amati02}. 
Comparing the values obtained when the full interval is split into
two sub-intervals, we find that the spectral index steepens from the
first to the second.  Likewise,
the $E^{\rm obs}_{\rm peak}$ value decreases dramatically from
$51.6^{+6.2}_{-5.5}$ keV to $20.6^{+2.7}_{-2.7}$ keV.

The overall emission characteristics of the burst are estimated from
the best-fit Band function, and summarized in Table
\ref{tbl:emission}.  In particular, the peak photon number and photon
energy fluxes in 1 s are
$(34.0 \pm 1.8)$ ph cm$^{-2}$ s$^{-1}$ and
$(1.68 \pm 0.11) \times 10^{-6}$ erg cm$^{-2}$ s$^{-1}$
in the 2--400 keV energy band.
\footnote{We compute the peak photon number flux using the ratio  2.672
of the flux of photons found in the 1 s time interval containing the
largest number of photons and the average flux of photons in the 5 s
time interval containing the  largest number of photons and that
brackets the 1 s time interval.  We compute the peak photon energy flux
in exactly the same way, except that  we use the ratio 2.782 of the
total photon energy flux (found by weighting each photon with its
energy and summing the energies) found in the 1 s time interval
containing the largest total photon energy and the average photon
energy flux in the 5 s time interval containing the largest total
photon energy and that brackets the 1 s time interval.}
  The energy fluence over the first 8.0 s of the burst was $S_X =1.36
\pm 0.05 \times 10^{-6}$ erg cm$^{-2}$ in the 2--30 keV band and
$S_\gamma = 2.17 \pm 0.15 \times 10^{-6}$ erg cm$^{-2}$ in the 30-400
keV energy band.  This gives a hardness ratio $S_X / S_\gamma$ of 0.63,
which is softer than many of the bursts seen by HETE-2 to date
\citep{barraud35} and makes this an ``X-ray-rich'' GRB.
\footnote{Throughout this paper, we define ``X-ray-rich'' GRBs and XRFs
as those events for which $\log [S_X(2-30~{\rm
keV})/S_\gamma(30-400~{\rm keV})] > -0.5$ and 0.0, respectively.}

\section{Discussion} \label{discussion}

\subsection{Burst Properties}

GRB 021211 was bright, and its duration was relatively short ($t_{50} =
0.86 \pm 0.07$~s and $t_{90} = 2.30 \pm 0.52$~s in the FREGATE 85--400
keV energy band; see Table \ref{tbl:temporal}) for a long-duration GRB
\citep{hurley92,lamb93,kouv93}.
The duration of the burst decreases with  increasing energy and the
spectrum of the burst evolves from hard to soft; both properties  are
typical of long-duration GRBs.  The burst is ``X-ray-rich,'' with
a soft spectrum having a peak energy of only $E_{\rm peak} \approx 45$
keV, and exhibits a soft tail, which may indicate that the X-ray
afterglow of the burst had already begun at this time.

\subsection{Optical Follow-Up Observations}

Previous to GRB 021211, only two GRB optical afterglows had been caught
in near-real time: the optical afterglow of GRB 990123, which was
caught by a single robotic telescope \citep{akerlof99}; and the optical
afterglow of GRB 021004 (another HETE-2 burst), which was caught by two
automated telescopes and a robotic telescope \citep{fox03a}.

The real-time localization of GRB 021211 by the WXM flight software,
which was rapidly followed by a refined WXM location and an SXC
location, both based on ground analysis, led to identification of the
optical afterglow in automated, filterless observations, using the
48-inch Schmidt telescope at Mt. Palomar, taken 20 minutes after the
start of the burst, and reported 53 minutes later %
\citep{gcn1731}.  

The prompt reporting of the location of GRB 021211 by the HETE-2 flight
software made possible an unprecedented number of successful near-real
time observations of the optical afterglow by three robotic telescopes:
RAPTOR \citep{gcn1757},
KAIT \citep{gcn1737},
and Super-LOTIS \citep{gcn1736}.  
The observations began 65 s, 108 s, and 143 s, respectively, after the
start of the burst, and revealed a light curve for the optical
afterglow that decreased rapidly during the first 12 minutes ($\sim
t^{-1.82 \pm 0.02}$), followed by a slower decline ($\sim t^{-0.82 \pm
0.11}$)
\citep{li03}. 
The latter decay rate is not atypical of the early-time behavior of GRB
afterglows
(see Figure 2 of Fox et al. 2003b).

The host galaxy of the burst was identified in HST images taken using
the ACS on 18 December UT (six days after the burst) and using the
NICMOS on 24 and 25 December UT (twelve and thirteen days after the
burst) %
\citep{gcn1781}. 
Soon thereafter, observations at the European Southern Observatory
using the Very Large Telescope showed that the redshift of the host
galaxy is $z = 1.006$
\citep{gcn1785}. 

\subsection{Implications for the Nature of GRB Jets}

GRB 021211 is the second burst for which an exceptionally bright,
rapidly decaying component of the optical afterglow has been seen, the
other being the famous event GRB 990123
\citep{briggs99,akerlof99}.
The rapidly decaying component in GRB 990123 has been interpreted as due
to emission from the reverse shock resulting from the the GRB jet
striking circumburst material %
\citep{sari99a,sari99b}.
The behavior of the bright, rapidly decaying component of the optical
afterglow of GRB 021211 also appears to be consistent with this
interpretation \citep{fox03b}.

\cite{zhang2003} and \cite{kumar2003} have shown that knowledge of both
the early-time behavior of the optical afterglow (which is thought to
be due to the reverse shock component), and the late-time behavior
(which is thought to be due to the forward shock component) provides a
diagnostic of whether or not the GRB jet is magnetic energy dominated. 
A detailed analysis by \cite{kumar2003} shows that the early-  and
late-time behavior of the optical afterglow of GRB 021211 requires that
the magnetic energy density in the jet exceed the kinetic energy in the
ejecta by a factor $\sim 1000$.  \cite{zhang2003} reach a similar
conclusion for GRB 990123 \citep{briggs99}, the only other GRB which is
known to have had an optical afterglow that was very bright at early
times \citep{akerlof99}.

\subsection{Implications for the Nature of ``Optically Dark'' GRBs}

Two explanations of why some GRBs are not detected optically, despite
early and deep follow-up observations, have been widely discussed: (1)
the optical afterglow is extinguished by dust in the host galaxy of the
burst (see, e.g., Reichart \& Price 2002) or (2) the GRB lies at very
high redshifts ($z > 5$), and the optical afterglow is absorbed by
neutral hydrogen in the host galaxy and in the intergalactic medium
along the line of sight from the burst to us (see, e.g., Lamb \&
Reichart 2000).  A third explanation has also been mentioned:  some
GRBs have optical afterglows that are very faint
\citep{fynbo2001,berger2002,lazzati2002}.

The optical follow-up observations of GRB 021211 made possible by
HETE-2 show that the optical afterglow of this burst was intrinsically
much fainter at late times than most afterglows observed previously. 
In particular, the optical afterglow of GRB 021211 faded from $\approx
14$ mag to $> 19$ mag within $\approx$ 20 minutes \citep{gcn1731}, and
was fainter than 23rd magnitude within one day
\citep{gcn1739,gcn1744,gcn1750,fox03b,pandey2003}.  The afterglow of
GRB 021211 was therefore more than three magnitudes fainter than the
afterglow of GRB 990123 at similar epochs \citep{akerlof99}.  It was
$\approx$ 2 magnitudes fainter than the faint optical afterglow of GRB
020124 \citep{berger2002} at 100 minutes after the burst, and remained
fainter at all subsequent epochs.  It was also fainter at one day than
the faint optical afterglows of GRB 980613 \citep{hjorth2002} and GRB
000630 \citep{fynbo2001}.  

Indeed, about 70\% of the searches conducted to date would have failed
to detect GRB 021211 at one day after the burst (see Berger et al., Figure 4;
Lazzati, Covino \& Ghisellini 2002, Figure 1; Fox et al. 2003b, Figure
2).  In addition, according to the statistics compiled by Reichart and
Yost (2003, Figure 1), 65\% (15/23) of GRBs for which optical
observations had been made with limiting magnitudes 20 $<$ R $<$ 23.5
at 18 hours after the burst were classified as ``optically dark.''  It
is therefore quite likely that the optical afterglow of GRB 021211
would not have been detected, were it not for the real-time
localization of the burst by the HETE-2 WXM, and the refined WXM and
SXC ground localizations that were rapidly disseminated, and would
therefore have been classified as an ``optically dark'' GRB.

Thus GRB 021211 shows that some fraction of burst afterglows are
``optically dark'' because their optical afterglows at times $>$ 1 hour
after the burst are very faint, and previously have often escaped
detection.  These bursts are ``optically dim'' rather than truly
``optically dark.'' Their existence and properties may lead to an
improved understanding of GRBs themselves, as well as burst
afterglows.  Their existence (as demonstrated by GRB 021211) also
suggests that the current sample of burst optical afterglows is not a
``fair sample,'' but is skewed toward bright afterglows.

It is interesting to compare the optical afterglow of GRB 021211 and
the optical afterglows of GRB 990123 \citep{briggs99} and GRB 021004
\citep{shirasaki2002}, the only other bursts for which successful real
time or near-real time optical afterglow observations have been made
\citep{akerlof99,fox03a}.  The afterglows of both GRB 021211 and GRB
990123 show a rapid decline during the first 20 minutes after the
burst, followed by a slower decline, approximately as $\sim t^{-1}$;
the latter is similar to the behavior of the afterglows of many GRBs at
early times.  Indeed, the temporal behaviors of the two afterglows are
almost carbon copies of each other, with the exception that the optical
afterglow of GRB 021211 is more than 3 magnitudes fainter than that of
GRB990123 at the same epoch \citep{akerlof99,fox03b}.  In contrast, the
optical afterglow of GRB 021004, another burst that was localized in
near-real time by HETE-2 \citep{shirasaki2002} and the only other burst
for which near-real time optical afterglow observations have been made
\citep{fox03a}, exhibited a very different behavior at early times:
the  afterglow faded only gradually in the first $\approx$ 2.5 hours
after the burst.  These results show that the optical afterglows of
GRBs exhibit a variety of behaviors at early times, and illustrate the
fact that HETE-2 is making it possible to explore the previously
unknown behavior of GRB afterglows in the ``gap'' in time from the end
of the burst to 3--20 hours later that existed during the {\it
Beppo}SAX era.

\subsection{Implications for Using GRBs as a Probe of Cosmology}

Confirmation that some GRBs have very bright optical afterglows at
early times, even if the afterglows are ``optically dim'' at late
times, has significant implications for using GRB afterglows as a probe
of cosmology and the early universe.  In particular, the bright early
phase of the optical afterglows of GRB 021211 (which lasted $\approx$
20 minutes) and GRB 990123 (which lasted $\approx$ 30 minutes), would
have lasted $\approx$ 1, 2, and 3.5 hours had these burst occurred at
redshifts $z$ = 5, 10, and 20, rather than at $z = 1.06$ and $z = 1.6$,
respectively.  This means that the spectral intensity of the afterglows
of GRBs at very high redshifts ($z > 5$) available at a fixed time
after the burst for NIR and infrared observations could be much greater
within the first few hours than has previously been assumed
\citep{lamb00,ciardi2000}.

\section{Conclusions} \label{conclusions}

GRB 021211 demonstrates that real-time localization of GRBs is critical
to acquiring observations which can help us understand both the transition
between the burst itself and its X-ray, optical, and radio afterglows,
and the distinct, early behavior of the burst afterglow, including the
interpretation that the bright, rapidly fading component is due to
emission from a reverse shock.

GRB 021211 also demonstrates that ``optically dark'' GRBs are due not
only to extinction of the optical afterglow by dust in the host galaxy,
and to absorption of the optical afterglow by neutral hydrogen in the
host galaxy and the IGM, if there are bursts at very high redshifts,
but also because the optical afterglows of some GRBs are much fainter
than those observed to date and are therefore not detected.  The
optical afterglows of these bursts should properly be called
``optically dim,'' rather than ``optically dark.''  However, GRB 021211
shows that even such ``optically dim'' bursts can have very bright
optical afterglows at times $<$ 20 minutes after the burst.  This
suggests that the UVOT onboard {\it Swift} may be able to detect and
localize a number of ``optically dark'' GRBs.  The confirmation that
some GRBs have very bright optical afterglows at early times, and that
even GRBs whose afterglows are ``optically dim'' at late times can have
bright afterglows at early times, also has significant implications for
using GRBs and their afterglows as a probe of cosmology and the early
universe.  It means that the spectral intensity of the afterglows of
GRBs at very high redshifts ($z > 5$) at a fixed time after the burst
could be much greater within the first few hours than has previously
been assumed.

\acknowledgments
\section*{Acknowledgments}

We would like to thank the anonymous referee for comments and
suggestions that materially improved the paper.  We would also like to
thank Scott Barthelmy for {\it sine qua non} GCN support.  The HETE-2
mission is supported in the US by NASA contract NASW-4690; in Japan, in
part by the Ministry of Education, Culture, Sports, Science, and
Technology Grant-in-Aid 13440063; and in France, by CNES contract
793-01-8479.  KH is grateful for HETE-2 support under Contract
MIT-SC-R-293291, for Ulysses support under JPL Contract 958056, and for
IPN support under NASA grant FDNAG5-11451.  G. Pizzichini acknowledges
support by the Italian Space Agency.

\clearpage

\begin{deluxetable}{lllcc}
\tablecaption{GRB 021211 Location History.
\label{tbl:location}}
\tablewidth{0pt}
\tablehead{
\colhead{Source} &
\colhead{$\alpha_{\rm J2000.0}$} & 
\colhead{$\delta_{\rm J2000.0}$} &
\colhead{Radius} & \colhead{Offset}
}
\startdata
WXM Alert        & 08 08 57.1   &  06 42 47    & 14\arcmin  & 1.1\arcmin \\
WXM Update       & 08 08 54.7   &  06 44 05    & 14\arcmin  & 1.4\arcmin \\
SXC Flight       & 08 09 07.0   &  06 43 48    & --         & 1.6\arcmin \\
WXM Ground       & 08 09 04.8   &  06 40 41    & 5\arcmin   & 3.2\arcmin \\
SXC Ground       & 08 09 00.0   &  06 44 20    & 2\arcmin   & 0.7\arcmin \\
Optical Transient
                 & 08 08 59.858 &  06 43 37.52 & 0.1\arcsec & -- \\
\enddata
\vskip -18pt
\tablecomments{The units of right ascension are hours, minutes, and
seconds; those of declination are degrees, arcminutes, and arcseconds.
The position error radius is for 90\% confidence and the offset is from
the Fox \& Price (2002) optical transient.}
\end{deluxetable}

\begin{deluxetable}{lccc}
\tablecaption{Temporal Properties of GRB 021211.
\label{tbl:temporal}}
\tablewidth{0pt}
\tablehead{
\colhead{Instrument} & \colhead{Energy}
& \colhead{$t_{50}$} & \colhead{$t_{90}$} \\
& \colhead{(keV)} & \colhead{(s)} & \colhead{(s)}
}
\startdata
HETE-2 SXC     & \GBC2--10\GBC    & 3.9  $\pm$ 0.9  &  9.6 $\pm$ 5.4     \\
&&& \\
HETE-2 WXM     & \GBC2--25\GBC    & 2.8  $\pm$ 0.2  &  8.0 $\pm$ 0.2     \\
             & \GBC2--10\GBC    & 3.1  $\pm$ 0.1  &  8.5 $\pm$ 0.5     \\
             & \GBC2--5\GBC\GBC & 4.8  $\pm$ 0.3  & 20.2 $\pm$ 1.1\GBC \\
             & \GBC5--10\GBC    & 2.3  $\pm$ 0.2  &  5.4 $\pm$ 0.5     \\
             & 10--25\GBC       & 1.9  $\pm$ 0.1  &  3.9 $\pm$ 0.2     \\
&&& \\
HETE-2 FREGATE & \GBC6--15\GBC    & 2.01 $\pm$ 0.08 & 4.92 $\pm$ 0.50    \\
             & 15--30\GBC       & 1.53 $\pm$ 0.07 & 4.33 $\pm$ 0.33    \\
             & 30--85\GBC       & 0.93 $\pm$ 0.05 & 2.41 $\pm$ 0.15    \\
             & 85--400          & 0.86 $\pm$ 0.07 & 2.30 $\pm$ 0.52    \\
\enddata
\vskip -18pt
\tablecomments{Errors are 1-$\sigma$; the SXC energy range is approximate.}
\end{deluxetable}

\begin{deluxetable}{lccc}
\tablecaption{Spectral Model Parameters for GRB 021211.
\label{tbl:spectrum}}
\tablewidth{0pt}
\tablehead{
\colhead{Parameter}
                    & \colhead{t = 0--2.3~s}
                    & \colhead{t = 2.3--8.0~s}
		    & \colhead{t = 0--8.0~s}
}
\startdata
Cutoff Power Law: & & & \\
\GBC Photon index $\alpha$ $(E^{-\alpha})$ &
 			$ \Gbc 0.533^{+0.091      }_{-0.096} $  &
 			$ \Gbc 1.071^{+0.165      }_{-0.171} $  &
 			$ \Gbc 0.930^{+0.077      }_{-0.079} $ \\
\GBC Peak energy ($E_{\rm peak}$)          &
 			$ \Gbc \GBC66.0^{+3.9\GBC\GBC}_{-3.6  }  $  &
 			$ \Gbc \GBC20.5^{+2.9\GBC\GBC}_{-2.5  }  $  &
 			$ \Gbc \GBC55.9^{+4.3\GBC\GBC}_{-3.7  }  $ \\
\GBC Cutoff energy ($E_{o}$ keV)               &
 			$ \Gbc \GBC45.0^{+5.3\GBC\GBC}_{-4.6  }  $  &
 			$ \Gbc \GBC22.1^{+6.2\GBC\GBC}_{-4.5  }  $  &
 			$ \Gbc \GBC52.3^{+7.3\GBC\GBC}_{-6.1  }  $ \\
\GBC Normalization (at 15 keV)             &
 			$ \Gbc 0.744^{+0.040\GBC   }_{-0.038 }  $  &
 			$ \Gbc 0.291^{+0.068\GBC   }_{-0.051 }  $  &
 			$ \Gbc 0.345^{+0.020\GBC   }_{-0.018 }  $ \\
Band Function: & & & \\
\GBC Index $\alpha$                        &
 			$    -0.242^{+0.178      }_{-0.161} $  &
 			$    -1.078^{+0.182      }_{-0.151} $  &
 			$    -0.805^{+0.112      }_{-0.105} $ \\
\GBC Index $\beta$                         &
 			$ \GBC-2.31^{+0.15\GBC   }_{-0.23 } $  &
 			$ \GBC < -2.98                      $  &
 			$ \GBC-2.37^{+0.18\GBC   }_{-0.31 } $ \\
\GBC Peak energy ($E^{\rm obs}_{\rm peak}$ keV)      &
 			$\GBC\GBC51.6^{+6.2\GBC    }_{-5.5  } $  &
 			$\GBC\GBC20.6^{+2.7\GBC    }_{-2.7  } $  &
 			$\GBC\GBC46.8^{+5.8\GBC    }_{-5.1  } $ \\
\GBC Normalization (at 15 keV)             &
 			$ \GBC0.911^{+0.131      }_{-0.098} $  &
 			$ \GBC0.287^{+0.074      }_{-0.046} $  &
 			$ \GBC0.389^{+0.042      }_{-0.034} $ \\
\enddata
\vskip -18pt
\tablecomments{Errors are for 90\% confidence.  The normalization
units are ph cm$^{-2}$ s$^{-1}$ keV$^{-1}$.}
\end{deluxetable}

\begin{deluxetable}{rccccc}
\tablecaption{Emission Properties of GRB 021211.
\label{tbl:emission}}
\tablewidth{0pt}
\tablehead{
\colhead{Energy} &
\colhead{Peak Photon Flux} &
\colhead{Photon Fluence} &
\colhead{Peak Energy Flux} & 
\colhead{Energy Fluence} \\
\colhead{(keV)} & \colhead{(ph cm$^{-2}$ s$^{-1}$)} & 
\colhead{(ph cm$^{-2}$)} & 
\colhead{($10^{-7}$ erg cm$^{-2}$ s$^{-1}$)} & \colhead{($10^{-7}$ erg cm$^{-2}$)}
}
\startdata
 \GBC\GBC2--10\GBC & $11.3 \pm 1.6$ & $55.6 \pm 4.9$ & 
$1.03 \pm 0.12$ & $4.41 \pm 0.33$ \\

\GBC\GBC2--25\GBC & $21.4 \pm 1.8$ & $83.4 \pm 5.4$ & 
$3.81 \pm 0.19$ & $11.6 \pm 0.5$ \\

\GBC\GBC2--30\GBC & $23.5 \pm 1.8$ & $88.1 \pm 5.4$ & 
$4.74 \pm 0.21$ & $13.6 \pm 0.5$ \\

 \GBC\GBC7--30\GBC & $15.4 \pm 0.67$ & $44.6 \pm 1.6$ & 
$4.17 \pm 0.17$ & $10.9 \pm 0.4$ \\

 \GBC30--400       & $10.5 \pm 0.44$ & $21.7 \pm 0.96$ & 
$12.2 \pm 0.89$ &  $21.7 \pm 1.5$ \\

 \GBC50--100       & $3.84 \pm 0.30$ & $8.32 \pm 0.40$ & 
$4.37 \pm 0.33$ & $9.13 \pm 0.47$ \\

 100--300          & $1.51 \pm 0.28$ & $2.64 \pm 0.40$ & 
$4.04 \pm 0.63$ & $5.86 \pm 1.03$ \\

  2--400          & $34.0  \pm 1.8$ & $109.8 \pm 5.5$ & 
$16.8 \pm 1.1$ & $35.4 \pm 1.6$ \\
\enddata
\vskip -18pt
\tablecomments{All of the quantities in this table are derived assuming
a Band function for the spectrum.  Errors are for 90\% confidence.}
\end{deluxetable}

\clearpage

\begin{figure}
\includegraphics[angle=270,scale=0.7]{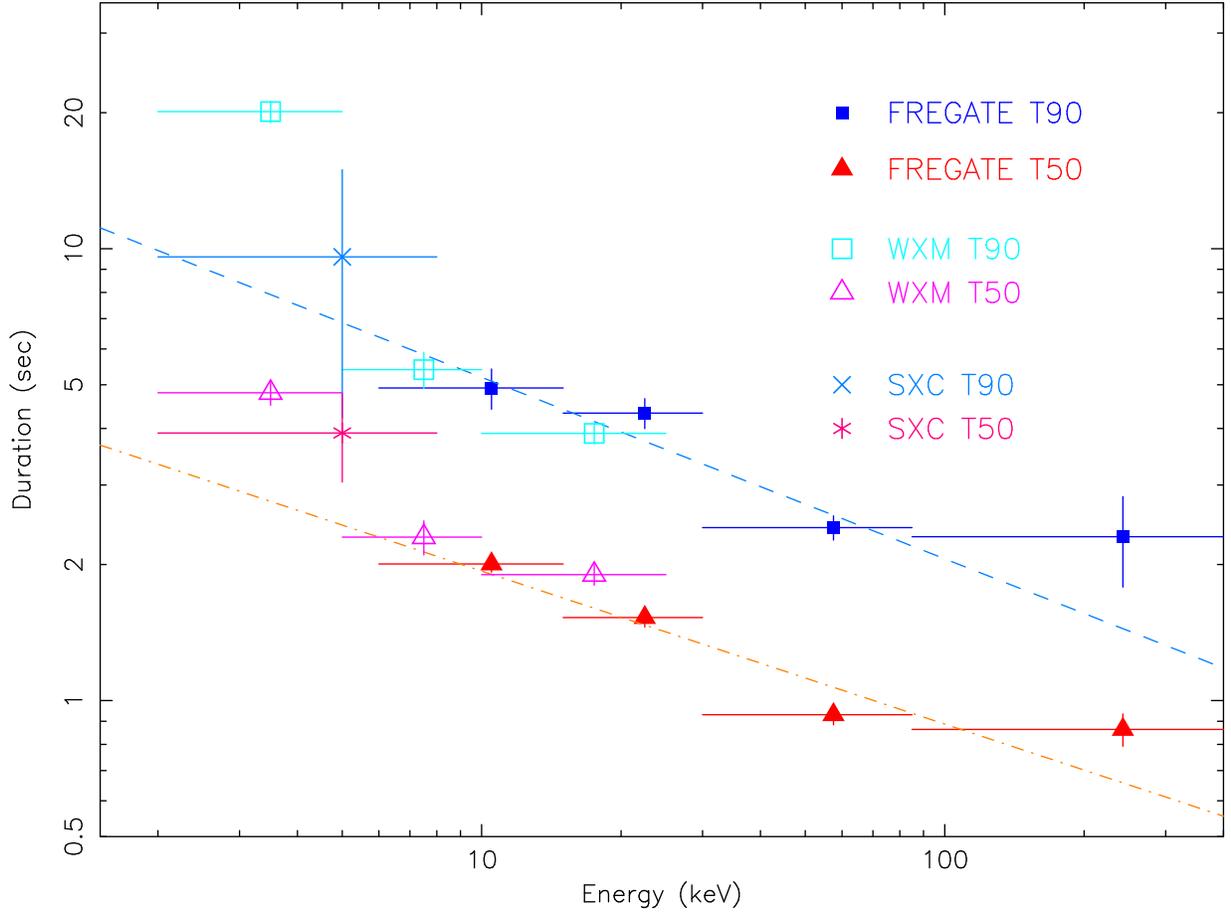}
\caption{Duration of GRB 021211 as a function of energy.
The $t_{50}$ and $t_{90}$ values for the FREGATE data increase
with decreasing energy as $E^{-0.34}$ and $E^{-0.40}$
(dashed lines), respectively.
The FREGATE data were binned into 6--15~keV, 15--30~keV, 30--85~keV,
and 85--300~keV bands; the WXM data were binned into 2--5~keV, 5--10~keV,
and 10--15~keV bands.  The SXC data were not subdivided, and the large
error bar is on the $t_{90}$ datum (see Table \ref{tbl:temporal}).
\label{fig:t5090}}
\end{figure}

\begin{figure}
\includegraphics[angle=270,scale=0.7]{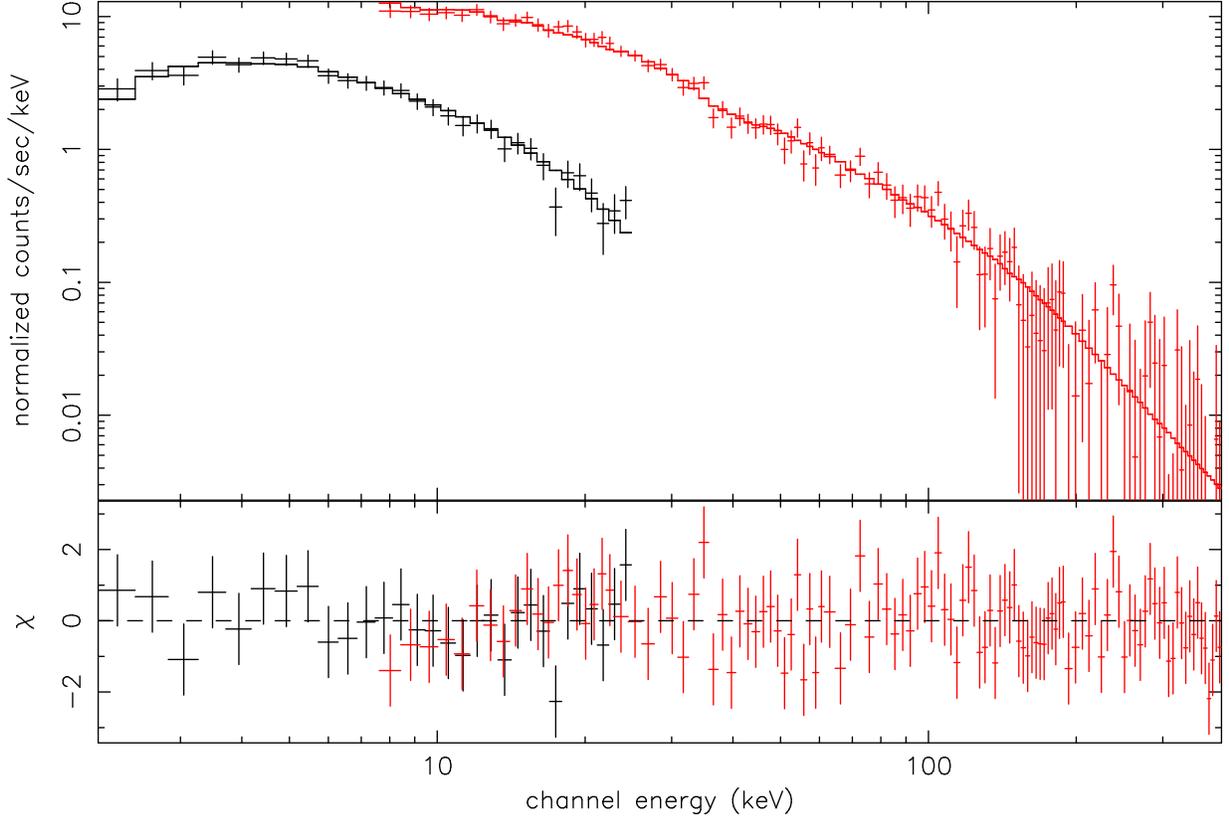}
\caption{Comparison of the observed and predicted spectrum of GRB
021211 in count space, integrated over the first 8.0 s of the burst. 
The upper panel compares the counts in the WXM energy loss channels
(lower energies) and the FREGATE energy loss channels (higher energies)
and those predicted by the best-fit Band function spectral model
($\alpha = -0.805^{+0.112}_{-0.105}, \beta = -2.37^{+0.18}_{-0.31},
E^{\rm obs}_{\rm peak} = 46.8^{+5.8}_{-5.1}$ keV); the lower panel
shows the residuals to the fit.
\label{fig:spec-cts}}
\end{figure}

\end{document}